%% file: isita08paper.tex
\documentclass[conference]{IEEEtran}
\usepackage{amsmath}
\usepackage{amsthm}
\usepackage{amssymb}
\usepackage{bm}
\usepackage{xspace}
\usepackage{xcolor}
\usepackage{graphicx}
\usepackage{psfrag}
\usepackage{url}
\usepackage{flushend}

\input{macros}

\title{Maximum Entropy Rate of Markov Sources\\
for Systems With Non-regular Constraints}

\author{\IEEEauthorblockN{G. B\"ocherer\IEEEauthorrefmark{1},
V.C. da Rocha Jr.\IEEEauthorrefmark{2},
C. Pimentel\IEEEauthorrefmark{2} and
R. Mathar\IEEEauthorrefmark{1}}
\IEEEauthorblockA{\IEEEauthorrefmark{1}Institute for Theoretical Information
Technology\\
RWTH Aachen University,
52056 Aachen, Germany\\ Email: \{boecherer,mathar\}@ti.rwth-aachen.de}
\IEEEauthorblockA{\IEEEauthorrefmark{2}
Communications Research Group - CODEC\\
Department of Electronics and Systems, P.O. Box 7800\\
Federal University of Pernambuco\\
50711-970 Recife PE, BRAZIL\\
E-mail: \{vcr,cecilio\}@ufpe.br
}}

\begin{document}

\maketitle

\input{abstract}

\input{introduction}

\input{dnc}

\input{sources}

\input{problem}

\input{result}

\input{conclusions}

\input{acknowledgement}

\bibliographystyle{IEEEtran}
\bibliography{confs-jrnls,Literatur}

\end{document}

%% file: macros.tex
\newtheorem{theorem}{Theorem}
\newtheorem{lemma}{Lemma}

\theoremstyle{definition}
\newtheorem{definition}{Definition}

\theoremstyle{remark}

\theoremstyle{remark} 
\newtheorem{exampleAux}{Example}
\newenvironment{example}[1] {
  \begin{exampleAux}\label{#1}
  }{ \hfill$\vartriangleleft$
  \end{exampleAux}
}
\newcommand{\safemath}[2]{\newcommand{#1}{\ensuremath{#2}\xspace}}

\safemath{\dncA}{\mathcal{A}} 
\safemath{\dncB}{\mathcal{B}}
\safemath{\dncC}{\mathcal{C}} 
\safemath{\dncD}{\mathcal{D}}
\safemath{\dncM}{\mathcal{M}}
\safemath{\setP}{\mathcal{P}}

\safemath{\mysetN}{\mathbb{N}} \safemath{\mysetC}{\mathbb{C}}
\safemath{\mysetR}{\mathbb{R}} \safemath{\mysetRP}{\mysetR^\oplus}

\safemath{\suppX}{\mathfrak{X}}
\safemath{\suppY}{\mathfrak{Y}}

\safemath{\emptystring}{\varepsilon}
\newcommand{\repart}[1]{\Re\left\lbrace #1\right\rbrace}
\safemath{\cop}{C}
\safemath{\comb}{\mathsf{C}}
\safemath{\cprob}{\mathsf{R}}
\safemath{\er}{\bar{H}}

\safemath{\emptychar}{\texttt{''}}
\safemath{\achar}{\texttt{t}}
\safemath{\bchar}{\texttt{u}}

\safemath{\aachar}{\achar\achar}
\safemath{\bbchar}{\bchar\bchar}
\safemath{\abchar}{\achar\bchar}
\safemath{\bachar}{\bchar\achar}

\safemath{\aabchar}{\achar\achar\bchar}
\safemath{\babchar}{\bchar\achar\bchar}

\newcommand{\weight}[1][]{\nu_{#1}}
\safemath{\wfunc}{\omega}
\safemath{\wset}{\Omega}
\safemath{\wmax}{\weight[\max]}
\safemath{\etamax}{\eta_{\max}}

\safemath{\erw}{H_{\weight}}
\safemath{\ern}{\overline{\mathrm{H}}}

\safemath{\bbranch}{\mathfrak{b}}

\safemath{\ppath}{\mathfrak{p}}
\safemath{\qpath}{\mathfrak{q}}
\safemath{\rpath}{\mathfrak{r}}

\safemath{\rnode}{\mathsf{r}}

\safemath{\vecx}{\mathbf{x}}
\safemath{\vecrho}{\pmb{\rho}}
\safemath{\matA}{\mathbf{A}}
\safemath{\matP}{\mathbf{P}}
\safemath{\matQ}{\mathbf{Q}}

\safemath{\eval}{\bigl.\bigr|}
\safemath{\iof}{\mathrm{i.o.}}
\safemath{\aev}{\mathrm{a.e.}}
\newcommand{\roc}{r.o.c.\@\xspace}

\DeclareMathOperator{\gf}{G}

\DeclareMathOperator{\probop}{P}
\DeclareMathOperator{\entop}{H}

\DeclareSymbolFont{BoldMath}{OML}{cmm}{b}{it}

\DeclareMathSymbol{\boa}{\mathalpha}{BoldMath}{'141}
\DeclareMathSymbol{\bob}{\mathalpha}{BoldMath}{'142}
\DeclareMathSymbol{\boc}{\mathalpha}{BoldMath}{'143}
\DeclareMathSymbol{\bod}{\mathalpha}{BoldMath}{'144}
\DeclareMathSymbol{\boe}{\mathalpha}{BoldMath}{'145}
\DeclareMathSymbol{\bog}{\mathalpha}{BoldMath}{'147}
\DeclareMathSymbol{\boh}{\mathalpha}{BoldMath}{'150}
\DeclareMathSymbol{\bom}{\mathalpha}{BoldMath}{'155}
\DeclareMathSymbol{\bon}{\mathalpha}{BoldMath}{'156}
\DeclareMathSymbol{\bop}{\mathalpha}{BoldMath}{'160}
\DeclareMathSymbol{\boq}{\mathalpha}{BoldMath}{'161}
\DeclareMathSymbol{\bor}{\mathalpha}{BoldMath}{'162}
\DeclareMathSymbol{\bos}{\mathalpha}{BoldMath}{'163}
\DeclareMathSymbol{\bot}{\mathalpha}{BoldMath}{'164}
\DeclareMathSymbol{\bou}{\mathalpha}{BoldMath}{'165}
\DeclareMathSymbol{\bov}{\mathalpha}{BoldMath}{'166}
\DeclareMathSymbol{\boxx}{\mathalpha}{BoldMath}{'170}
\DeclareMathSymbol{\boy}{\mathalpha}{BoldMath}{'171}
\DeclareMathSymbol{\boz}{\mathalpha}{BoldMath}{'172}
\DeclareMathSymbol{\bomu}{\mathalpha}{BoldMath}{'026}

\DeclareMathSymbol{\boA}{\mathalpha}{BoldMath}{'101}
\DeclareMathSymbol{\boB}{\mathalpha}{BoldMath}{'102}
\DeclareMathSymbol{\boC}{\mathalpha}{BoldMath}{'103}
\DeclareMathSymbol{\boD}{\mathalpha}{BoldMath}{'104}
\DeclareMathSymbol{\boE}{\mathalpha}{BoldMath}{'105}
\DeclareMathSymbol{\boF}{\mathalpha}{BoldMath}{'106}
\DeclareMathSymbol{\boG}{\mathalpha}{BoldMath}{'107}
\DeclareMathSymbol{\boH}{\mathalpha}{BoldMath}{'110}
\DeclareMathSymbol{\boI}{\mathalpha}{BoldMath}{'111}
\DeclareMathSymbol{\boJ}{\mathalpha}{BoldMath}{'112}
\DeclareMathSymbol{\boK}{\mathalpha}{BoldMath}{'113}
\DeclareMathSymbol{\boM}{\mathalpha}{BoldMath}{'115}
\DeclareMathSymbol{\boO}{\mathalpha}{BoldMath}{'117}
\DeclareMathSymbol{\boP}{\mathalpha}{BoldMath}{'120}
\DeclareMathSymbol{\boQ}{\mathalpha}{BoldMath}{'121}
\DeclareMathSymbol{\boR}{\mathalpha}{BoldMath}{'122}
\DeclareMathSymbol{\boS}{\mathalpha}{BoldMath}{'123}
\DeclareMathSymbol{\boT}{\mathalpha}{BoldMath}{'124}
\DeclareMathSymbol{\boU}{\mathalpha}{BoldMath}{'125}
\DeclareMathSymbol{\boV}{\mathalpha}{BoldMath}{'126}
\DeclareMathSymbol{\boW}{\mathalpha}{BoldMath}{'127}
\DeclareMathSymbol{\boX}{\mathalpha}{BoldMath}{'130}
\DeclareMathSymbol{\boY}{\mathalpha}{BoldMath}{'131}
\DeclareMathSymbol{\boZ}{\mathalpha}{BoldMath}{'132}

\safemath{\rvecA}{\boA}
\safemath{\rvecB}{\boB}
\safemath{\rvecC}{\boC}
\safemath{\rvecD}{\boD}
\safemath{\rvecE}{\boE}
\safemath{\rvecF}{\boF}
\safemath{\rvecG}{\boG}
\safemath{\rvecH}{\boH}
\safemath{\rvecI}{\boI}
\safemath{\rvecJ}{\boJ}
\safemath{\rvecK}{\boK}
\safemath{\rvecL}{\boL}
\safemath{\rvecM}{\boM}
\safemath{\rvecN}{\boN}
\safemath{\rvecO}{\boO}
\safemath{\rvecP}{\boP}
\safemath{\rvecQ}{\boQ}
\safemath{\rvecR}{\boR}
\safemath{\rvecS}{\boS}
\safemath{\rvecT}{\boT}
\safemath{\rvecU}{\boU}
\safemath{\rvecV}{\boV}
\safemath{\rvecW}{\boW}
\safemath{\rvecX}{\boX}
\safemath{\rvecY}{\boY}
\safemath{\rvecZ}{\boZ}
\safemath{\rvecZero}{\bZero}

\safemath{\rvecXl}{\rvecX^{(l)}}
\safemath{\suppXl}{\suppX^{(l)}}
\safemath{\vecxl}{\vecx^{(l)}}

%% file: abstract.tex
\begin{abstract}

Using the concept of discrete noiseless channels, it was shown by 
Shannon in \emph{A Mathematical Theory of Communication} that
the ultimate performance of an encoder for a constrained system is limited by
the combinatorial capacity of the system if the constraints define a regular
language.  In the present
work, it is shown that this is not an inherent property of regularity but holds
in general. To show this, constrained systems are described by generating
functions and random walks on trees.
\end{abstract}

%% file: introduction.tex
\section{INTRODUCTION}
A constrained system allows the transmission of input sequences  of weighted
symbols that fulfill certain constraints on the symbol
constellations. Constrained systems have been of recent interest, 
e.g., in the context of storage systems \cite{Immink2004}. A natural
question is how to efficiently encode a random source such that it becomes a
valid input for a constrained system \cite{Marcus2001}. Furthermore, it is of
interest to determine the ultimate performance of such an encoder. This leads to
the notion of the capacity of constrained systems.

\emph{Previous work:} Shannon~\cite{Shannon1948} investigated the capacity of constrained systems
within the framework of the \emph{discrete noiseless channel}
(DNC). For the case where the constraints form a regular language
\cite{Sipser2006}, it was stated in ~\cite[Theorem 8]{Shannon1948}
 that the maximum entropy rate \cprob of a valid input process is equal to the
\emph{combinatorial capacity} $\comb$, which is defined as
\begin{align}
\comb = \limsup_{\weight\rightarrow\infty}\frac{\ln
N(\weight)}{\weight}\label{eq:capacityOrig}
\end{align}
where $\weight$ denotes the length of the sequences and $N(\weight)$ denotes the number
of distinct sequences of length $\weight$ that are accepted by the considered
DNC. Here and hereafter, $\ln$ denotes the
natural logarithm. A detailed proof of the equality between \cprob and \comb was
recently given in \cite{Khandekar2000}. This proof is heavily based on the
regularity of the constraints.
However, it is not clear whether this equality
is an inherent property of regular languages or whether it holds in general. It should be
noted that sequences with non-regular constraints have been of research interest
recently, e.g., in \cite{Milenkovic2007}. An early treatment of DNCs can be
found in \cite{Csiszar1969}.

\emph{Contributions:} In this paper, we use the framework of \emph{general
DNCs} as introduced in \cite{Boecherer2007} to show the following.
If the set of valid input sequences for a constrained system can be
generated by a Markov process, then the maximum entropy rate of such a
process is given by the combinatorial capacity of the system, irrespective of
whether the constraints are regular or not. Our result can be seen as a
generalization of Shannon's result \cite[Theorem 8]{Shannon1948} to general
DNCs and in particular non-regular DNCs. Furthermore, since our derivations also
apply for the regular case, they also serve as a new way to derive
\cite[Theorem 8]{Shannon1948}.

The remainder of the paper is organized as follows. In Section~\ref{sec:dnc},
we present the framework of general DNCs and the calculation of combinatorial
capacities by generating functions as introduced in \cite{Boecherer2007}. We
then define in Section~\ref{sec:sources} Markovian input processes and entropy
rates for general DNCs. In Section~\ref{sec:problem}, we define the maximum
entropy rate \cprob of general DNCs and for sake of illustration, we show for
two simple examples that \cprob is equal to the combinatorial capacity \comb.
Finally, in Section~\ref{sec:result}  we prove that $\cprob=\comb$ holds for
general DNCs.

%% file: dnc.tex
\section{DISCRETE NOISELESS CHANNELS}\label{sec:dnc}

To calculate the combinatorial capacity of general DNCs,
we interpret generating functions as functions on the complex plane and
investigate their convergence behavior. This approach, mostly referred to as
\emph{analytic combinatorics}, is discussed in detail in \cite{Flajolet2008}. We
consider a more general case since we allow non-integer valued
symbol weights. In order to handle this situation, we use general Dirichlet's
series \cite{Hardy1915} instead of Taylor series as generating functions.
\subsection{Definitions and Notation}

Our definition of DNCs as presented next mainly follows the one given in
\cite{Boecherer2007}. 

\begin{definition}
  A DNC $\dncA=(A,\wfunc)$ consists of a countable set $A$ of strings
  accepted by the channel and an associated weight function $\wfunc\colon
  A\rightarrow\mysetRP$ (\mysetRP denotes the nonnegative real numbers)
  with the following property. If $a,b\in A$ and $ab\in A$
  ($ab$ denotes the concatenation of $a$ and $b$),
then $\wfunc(ab)=\wfunc(a)+\wfunc(b)$.
  By convention, the empty string $\emptystring$ is always an element of $A$ and
the weight of $\emptystring$ is equal to zero, i.e., $w(\emptystring)=0$.
\end{definition}
\begin{definition}\label{def:generatingFunction}
  Let $\dncA=(A,\wfunc)$ represent a DNC. We define the \emph{generating
    function} of $\dncA$ by
  \begin{align}
    \gf_\dncA(s)&=\sum\limits_{a\in A}e^{-\wfunc(a)s},\qquad s\in\mysetC
  \end{align}
where $\mysetC$ denotes the set of complex numbers.
\end{definition}
Let  $\wset$ denote the set of distinct string weights of elements in $A$. We
order and index the set $\wset$ such that
$\wset=\lbrace \weight[k]\rbrace_{k=1}^{\infty}$ with
\mbox{$\weight[1]<\weight[2]<\dotsb$}. For every $\weight[k]\in\wset$,
$N(\weight[k])$
denotes the number of distinct strings of weight $\weight[k]$ that are accepted
by the channel. We can now write the generating function as
\begin{align}
  \gf_\dncA(s)=\sum\limits_{k=1}^\infty N(\weight[k])e^{-\weight[k]s}.
\end{align}
Since the coefficients $N(\weight[k])$ result from an enumeration, they are all
nonnegative. The combinatorial capacity of a DNC as defined in 
\eqref{eq:capacityOrig} can now be written as
\begin{align}
    \comb=\limsup_{k\rightarrow\infty}\dfrac{\ln
N(\weight[k])}{\weight[k]}.\label{eq:capacity}
\end{align}

\subsection{DNCs of Interest}
Throughout this paper, we restrict our attention to DNCs where the ordered set
of string weights $\lbrace \weight[k]\rbrace_{k=1}^{\infty}$ is \emph{not too
dense}, that is, there exists some constant $L\geq 0$ and some constant $K\geq
0$ such that for any integer $n\geq 0$
\begin{align}
  \max_{\weight[k]<n} k\leq Ln^K\label{eq:notTooDense}.
\end{align}
Otherwise, the number of possible string weights in the interval $[n,n+1]$
increases exponentially with $n$, in which case the  definition of combinatorial
capacity given in \eqref{eq:capacity} is not appropriate. This is illustrated in
the following example.
\begin{example}{ex:tooDense}
  Let $N(\weight[k])$ denote the coefficients of the generating function of
  some DNC. Assume $N(\weight[k])=1$ for all $k\in\mysetN$ and assume further
  \begin{align}
    \max_{\weight[k]<n}k=\left\lceil R^n\right\rceil
  \end{align}
  for some $R>1$. According to \eqref{eq:capacity}, the
  capacity of the DNC is then equal to zero because of $\ln N(\weight[k])=0$ for
  all $k\in\mysetN$. However, the channel accepts $R^n$ distinct
  strings of weight smaller than $n$. The average amount of data per
  string weight that we can transmit over the channel is thus
  lower-bounded by $\ln R^n/n=\ln R$, which is according to the assumption
  greater than zero.
\end{example}
For a DNC $\dncA=(A,\wfunc)$ where $A$ is generated over a finite set of
symbols, the restriction \eqref{eq:notTooDense} is automatically fulfilled
\cite[Appendix A]{Khandekar2000}, implying that virtually any constrained system of
practical
interest fulfills \eqref{eq:notTooDense}. Not too dense sequences have another
interesting property, which we will need in our later derivations. We state it
in the following lemma.
\begin{lemma}\label{lem:notTooDenseConvergence}
 If a series $\{a_k\}_{k=1}^\infty$ is not too dense and if $0\leq x<1$, then
the series $\sum_{k=1}^\infty x^{a_k}$ converges.
\end{lemma}
 See \cite[Appendix A]{Khandekar2000} for a proof of this lemma.
\subsection{Calculating the Capacity}
For a DNC of interest, we want to calculate the 
combinatorial capacity as given in \eqref{eq:capacity}.
An explicit formula for regular DNCs was provided in 
\cite[Theorem 1]{Shannon1948}. A detailed derivation of this formula for DNCs
with regular constraints and non-integer valued symbol weights can 
be found in \cite{Khandekar2000}. In \cite{Boecherer2007}, it was shown that
the combinatorial capacity \eqref{eq:capacity} is determined by the region of
convergence (\roc) of the corresponding generating function for any DNC with the
set of possible string weights $\lbrace \weight[k]\rbrace_{k=1}^{\infty}$ being
not too dense. We restate this theorem here.
\begin{theorem}\label{theo:capacityConvergence}
Let $\dncA=(A,\wfunc)$ be a DNC with the generating function
 $\gf_\dncA(s)$. The combinatorial capacity $\comb$ of \dncA is given by $\comb=Q$ where $\repart{s}>Q$
($\repart{s}$ denotes the real part of $s$) is the \roc of $\gf_\dncA(s)$, that is,
\begin{align}
C = \limsup_{k\rightarrow\infty}\dfrac{\ln N(\weight[k])}{\weight[k]}=Q.
\end{align}
\end{theorem}
Theorem~\ref{theo:capacityConvergence} applies for general DNCs with possibly
non-integer valued symbol weights and arbitrary constraints on the symbol
constellations. It can be interpreted as the general form of the
\emph{Exponential Growth Formula}. In \cite[Theorem IV.7]{Flajolet2008},
the Exponential Growth Formula was stated for DNCs with integer valued weights
and arbitrary constraints. The latter version of the Exponential Growth 
Formula was used in \cite{Milenkovic2007} to calculate the combinatorial
capacity \eqref{eq:capacity} of a non-regular DNC with integer
valued symbol weights.

%% file: sources.tex
\section{INPUT SOURCES FOR DNCs}\label{sec:sources}
\begin{figure}
    \begin{center}
	\psfrag{a}{$\achar$}
	\psfrag{b}{$\bchar$}
	\psfrag{ab}{$\abchar$}
	\psfrag{E}{$\emptystring$}
	\psfrag{0}{$0$}
	\psfrag{1}{$1$}
	\psfrag{2}{$2$}
	\psfrag{w}{$\weight$}
	\psfrag{(a)}{i.}
	\psfrag{(b)}{ii.}
      \includegraphics[width=\linewidth]{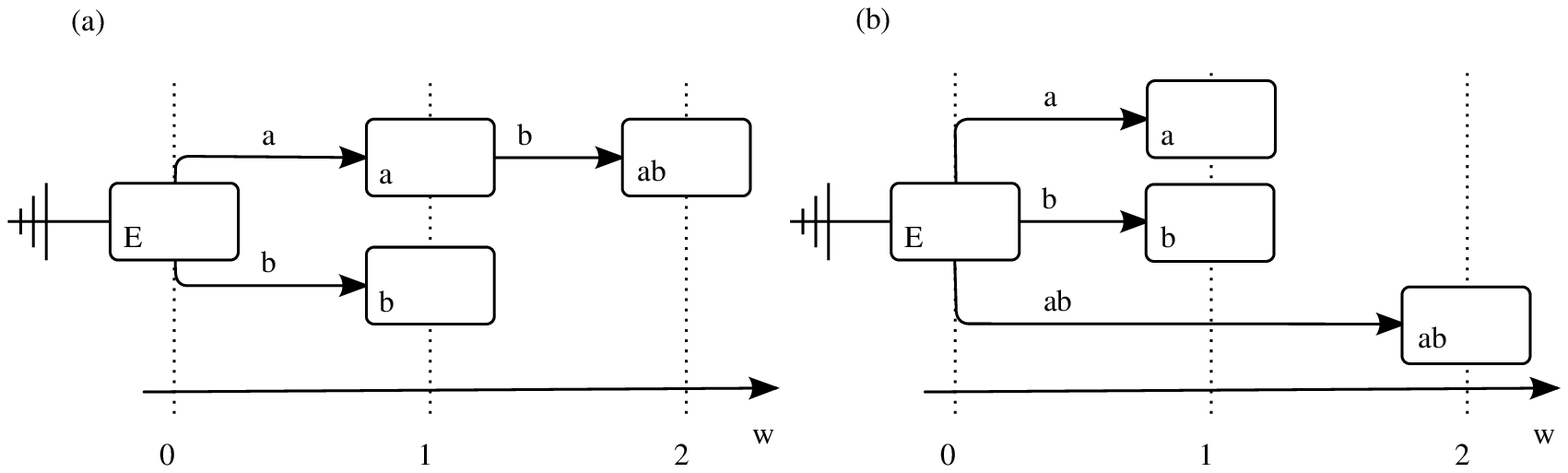}
    \end{center}
    \caption{
Two different representations of the DNC $\dncA=(A,w)$ by a tree.
The DNC $\dncA$ is given by $A=\{\emptystring,\achar,\bchar,\abchar\}$ with
$w(\achar)=w(\bchar)=1$.}
\label{fig:nonUniqueTree}
\end{figure}
The purpose of this section is to define Markovian input processes and
the corresponding entropy rates for general DNCs. First, we represent the set of
strings that are accepted by a general DNC by a tree and second, we define a
Markovian input process as a walk on this tree and give a formula for its
entropy rate. We postpone the problem of finding the maximum entropy rate to the
next section.
\subsection{Representing DNCs by Trees}
We represent a DNC $\dncA=(A,\wfunc)$ by a tree $T_\dncA$ consisting of a root,
labelled and weighted branches, and paths resulting from the concatenation of
branches. We restrict our considerations to paths that start at the root. For
each such path, we display its label at the corresponding end node. We do not
allow distinct paths to have the same label. A DNC $\dncA$ is represented by
a tree $T_\dncA$ if there is a one-to-one mapping from $A$ to the path labels.
Note that only the \emph{set of paths} in $T_\dncA$ is uniquely determined by
this mapping, but not how these paths are formed by branches. See
Figure~\ref{fig:nonUniqueTree} for an example of this ambiguity. In this
figure, a branch is represented by an arrow, its weight by the distance
between start and end node, and its label is written above the arrow. Notice that the set of
paths represented by the node labels displayed in the rectangles is
the same for the tree in Figure~\ref{fig:nonUniqueTree}~i and the tree in
Figure~\ref{fig:nonUniqueTree}~ii. 
The DNC has a finite set $A$ of accepted sequences, therefore, the tree
representations are finite. 
However, DNCs of non-zero combinatorial capacity have infinite sets of accepted strings and as a
consequence also infinite tree representations.
Surprisingly, we will see in the following that although the tree
representation of a DNC is not unique, as long as it allows the definition of a
Markov input source, the maximum entropy rate of this source will not
depend on the chosen tree representation.
\subsection{Markovian Input Sources}
For a DNC $\dncA=(A,\wfunc)$, we assume that every branch in the tree
representation
$T_\dncA$ has subsequent branches. We can then define an input source by a
Markov process $X=\lbrace X_l\rbrace_{l=1}^\infty$, where $X_l$ chooses randomly
among the branches that start at the end node of the realization of $X_{l-1}$.
Every realization of $\rvecXl=(X_1,\dotsc,X_l)$ is thus a path in $T_\dncA$
starting at the root and consisting of $l$ branches. The support of $\rvecXl$ is
given by the set of all such paths $\vecxl$  and we denote it by $\suppXl$.
Note that for $\dncA=(A,\wfunc)$, we have
\begin{align}
A=\bigcup\limits_{l=1}^\infty \suppXl.\label{eq:suppXlA}
\end{align}
Whenever it follows directly from the context, we omit for simplicity the
superscript $l$ and write $\vecx$ instead of $\vecxl$. For all
$\vecx\in\suppXl$, we have for the probability mass function (PMF)
$p_{\rvecXl}$ of $\rvecXl$
\begin{align}
p_{\rvecXl}(\vecx)&=\probop[X_1=x_1]\prod\limits_{i=2}^l
\probop[X_i=x_i|X_{i-1}=x_{i-1}].
\end{align}
We conclude that the existence of a tree representation $T_\dncA$ where each
branch has subsequent branches is equivalent to the existence of a Markovian
input source for $\dncA$. Note that Regular DNCs can be
represented by \emph{finite state machines} (FSMs) \cite{Sipser2006} and the
tree representation can be obtained from the corresponding
FSM. The resulting tree representation then has automatically the property that
each branch has subsequent branches.

Following \cite{Shannon1948},\cite{Khandekar2000}, the \emph{entropy rate}
$\ern$ of $X$ is given by
\begin{align}
\ern(X)=\limsup_{l\rightarrow\infty}\frac{\entop(\rvecXl)}{L_l}
\label{eq:entropyRate}
\end{align}
where $L_l$ is equal to the average weight of all $\vecx\in\suppXl$ with
respect to (w.r.t.) the PMF of $\rvecXl$ and where $\entop(\rvecXl)$ denotes the
entropy of $\rvecXl$ in nats.

%% file: problem.tex
\section{PROBLEM STATEMENT}\label{sec:problem}
We now come to the key topic of this paper: the maximization of the entropy
rate of input processes for general DNCs.
\subsection{Maximum Entropy Rate}

\begin{definition}\label{def:probabilisticCapacity}
 We define the \emph{maximum entropy rate} $\cprob$ of a DNC by
\begin{align}
 \cprob=\max_{X}\ern(X).\label{eq:probabilisticCapacity}
\end{align}
where the maximum is taken over all Markovian processes $X$ that generate 
valid input sequences for the DNC.
\end{definition}
Note that in \cite{Khandekar2000}, the term \emph{probabilistic capacity} was
used instead of maximum entropy rate. However, we prefer the latter term.

The entropy rate $\ern(X)$ is maximized, if each term of the sequence on the
right hand side of \eqref{eq:entropyRate} is maximized. For each $l$, the
maximum entropy per average branch weight
\begin{align}
 R_l&=\max_{p_{\rvecXl}}\frac{\entop(\rvecXl)}{L_l}
\label{eq:maximumEntropyRate}
\end{align}
is given by the greatest positive real solution of the equation
\begin{align}
\sum\limits_{\vecx\in\suppX^{(l)}}e^{-\wfunc(\vecx)s}=1.
\label{eq:maximumEntropie}
\end{align}
In addition, for all $\vecx\in\suppXl$, the PMF of $\rvecXl$ that achieves this
rate is uniquely given by
\begin{align}
 q_{\rvecXl}(\vecx)=e^{-\wfunc(\vecx)R_l}.
\end{align}
These two properties of $R_l$ were derived by using Lagrange Multipliers
in \cite{Marcus1957} and they were independently derived in \cite{Krause1962} by
using the bound
$\ln x \leq x-1$. We offer an alternative proof by applying the information
inequality \cite{Cover2006}, which states for the Kullback Leibler Distance
$D(\cdot\Vert\cdot)$ of two PMFs $p$ and $q$ that
\begin{align}
 D(p\Vert q)\geq 0
\end{align}
with equality if and only if $p=q$. We thus have
\begin{align}
 0&\geq -D(p_{\rvecXl}\Vert q_{\rvecXl})\\
&=\sum\limits_{\vecx\in\suppXl}p_{\rvecXl}(\vecx)
\ln\frac{q_{\rvecXl}(\vecx)}{p_{\rvecXl}(\vecx)}\\
&=H(\rvecXl)-R_l L_l
\end{align}
which implies
\begin{align}
\frac{H(\rvecXl)}{L_l}\leq R_l
\end{align}
with equality if and only if $p_{\rvecXl}=q_{\rvecXl}$. Combining
\eqref{eq:entropyRate}, \eqref{eq:probabilisticCapacity}, and
\eqref{eq:maximumEntropyRate}, we have
\begin{align}
\cprob &=\limsup_{l\rightarrow\infty}R_l
=\limsup_{l\rightarrow\infty}\max_{p_{\rvecXl}}\frac{\entop(\rvecXl)}{L_l}.
\label{eq:probabilisticCapacity2}
\end{align}
The form on the right hand side of \eqref{eq:probabilisticCapacity2} allows
us to compare the maximum entropy rate of a DNC to its combinatorial capacity
as given in \eqref{eq:capacity}. We illustrate this in the
following by two simple examples. 
\begin{example}{ex:equal}
Let $\dncA=(A,\wfunc)$ represent a DNC that accepts all binary input sequences.
The set $A$ is thus given by $A=\{0,1\}^\star$ where $^\star$ denotes the
regular operation \emph{star} \cite{Sipser2006}. We assume the symbol weights
$\wfunc(0)=\wfunc(1)=1$. The combinatorial capacity is given by
\begin{align}
\comb &= \limsup_{k\rightarrow\infty}\frac{\ln N(\weight[k])}{\weight[k]}\\
&= \limsup_{k\rightarrow\infty}\frac{\ln 2^k}{k}.\label{eq:capacityEqual}
\end{align}
To calculate the maximum entropy rate of \dncA, we note that for each
$\vecx\in\suppXl$, we have $\wfunc(\vecx)=l$ and in addition, the cardinality
of $\suppXl$ is given by $|\suppXl|=2^l$. The average weight $L_l$ of $\vecxl$
is thus given by $L_l=l$ and maximizing the entropy rate reduces to maximizing
the entropy of $\rvecXl$. The maximum entropy of $\rvecXl$ is given by 
$\max_{p_{\rvecXl}}\entop(\rvecXl)=\ln |\suppXl|$, see \cite{Cover2006}. All
together we have
\begin{align}
\cprob
&= \limsup_{l\rightarrow\infty}\max_{p_{\rvecXl}}\frac{\entop(\rvecXl)}{L_l}\\
& = \limsup_{l\rightarrow\infty}\frac{\max_{p_{\rvecXl}}\entop(\rvecXl)}{l}\\
& = \limsup_{l\rightarrow\infty}\frac{\ln |\suppXl|}{l}\\
& = \limsup_{l\rightarrow\infty}\frac{\ln 2^l}{l}.\label{eq:maximumRateEqual}
\end{align}
We see from \eqref{eq:capacityEqual} and \eqref{eq:maximumRateEqual} that the
maximum entropy rate of \dncA is equal to the combinatorial capacity, that is,
$\cprob=\comb$.
\end{example}
\begin{example}{ex:unequal} As in Example~\ref{ex:equal}, we consider a DNC
$\dncA=(A,\wfunc)$ that accepts all binary input sequences. However, we assume
the symbol weights $\wfunc(0)=1$ and $\wfunc(1)=2$. To show that
$\comb=\cprob$ also holds in this case, we have to explicitly calculate
$\comb$ and $\cprob$. To show equality by comparison as we did by
\eqref{eq:capacityEqual} and \eqref{eq:maximumRateEqual} in the previous
example is no longer possible. To calculate the combinatorial capacity, we
write the generating function of \dncA as
\begin{align}
\gf_\dncA(s)=\sum\limits_{m=0}^\infty (e^{-1s}+e^{-2s})^m.
\end{align}
The series converges if $\repart{e^{-1s}+e^{-2s}}<1$, therefore, the
combinatorial capacity $\comb$ is by Theorem~\ref{theo:capacityConvergence}
given by the smallest positive real solution of
\begin{align}
e^{-1s}+e^{-2s}=1.\label{eq:capacityUnequal}
\end{align}
Let $Y$ denote a random variable with support $\{0,1\}$, and the 
associated weights $\wfunc(0)=1$ and $\wfunc(1)=2$. In addition, let $L$ denote
the average weight of $Y$. The maximum entropy rate of \dncA can then be
calculated as
\begin{align}
\cprob
&=\limsup_{l\rightarrow\infty}\max_{p_{\rvecXl}}\frac{\entop(\rvecXl)}{L_l}\\
&=\limsup_{l\rightarrow\infty}\max_{p_Y}\frac{l\entop(Y)}{lL}\\
&=\max_{p_Y}\frac{\entop(Y)}{L}.
\end{align}
By \eqref{eq:maximumEntropie}, it follows from the last line that $\cprob$ is
also given by \eqref{eq:capacityUnequal}, thus $\cprob=\comb$.
\end{example}

%% file: result.tex
\section{MAIN RESULT}\label{sec:result}
Based on the concepts introduced in the previous sections, we can now state our
main result.
\begin{theorem}
\label{theo:ftdnc}
 If the set of valid input sequences of a DNC  $\dncA=(A,\wfunc)$ can be generated by a Markov
process (or equivalently, if the DNC can be represented by a tree where each
branch has a subsequent branch), then the maximum entropy rate \cprob of \dncA
is equal to its combinatorial capacity \comb, that is,
\begin{align}
\limsup_{k\rightarrow\infty}\dfrac{\ln
N(\weight[k])}{\weight[k]}=\limsup_{l\rightarrow\infty}\max_{p_{\rvecXl}}\frac{
\entop(\rvecXl)}{L_l}.
\end{align}
\end{theorem}
We will prove this equality in the following. Although equality was shown in
\cite{Khandekar2000} for regular DNCs, to the best of our knowledge nobody has
addressed the non-regular case until now.
\begin{proof}[Proof of Theorem~\ref{theo:ftdnc}]
 To proof the theorem, we show that the region of
convergence of the generating function $\gf_\dncA(s)$ is given by
$\repart{s}>\cprob$. The theorem then follows because
of Theorem~\ref{theo:capacityConvergence}.

The maximum entropy rate $\cprob$ is given by \eqref{eq:probabilisticCapacity2},
which is equivalent to the following. For every $\epsilon>0$, it holds that
\begin{align}
 R_l&< \cprob+\epsilon\quad\text{almost everywhere }(\aev)\label{eq:boundAev}\\
&\text{and}\nonumber\\
R_l&> \cprob-\epsilon\quad\text{infinitely often }(\iof)
\end{align}
with respect to $l\in\mysetN$ (the set of natural numbers). Since $R_l$ is given
by
\eqref{eq:maximumEntropie}, this implies further
\begin{align}
 \sum\limits_{\vecx\in\suppX^{(l)}}e^{-\wfunc(\vecx)[\cprob+\epsilon]}&<
\sum\limits_{\vecx\in\suppX^{(l)}}e^{-\wfunc(\vecx)R_l}=1\quad\aev
\label{eq:probabilitySumBoundAev} \\
&\text{and}\nonumber\\
 \sum\limits_{\vecx\in\suppX^{(l)}}e^{-\wfunc(\vecx)[\cprob-\epsilon]}&>
\sum\limits_{\vecx\in\suppX^{(l)}}e^{-\wfunc(\vecx)R_l}=1\quad\iof
\label{eq:probabilitySumBoundIof}
\end{align}
Because of \eqref{eq:suppXlA}, we can write the generating function as
\begin{align}
 \gf_\dncA(s)&=\sum\limits_{a\in A}e^{-\wfunc(a)s}\\
&=\lim_{n\rightarrow\infty}
\sum\limits_{l=1}
^n\sum\limits_{\vecx\in\suppX^ {(l)}} e^ {-\wfunc(\vecx)s}
\end{align}
and we can use \eqref{eq:probabilitySumBoundAev} and
\eqref{eq:probabilitySumBoundIof} to give bounds on $\gf_\dncA(s)$ around
$s=\cprob$. It follows directly from \eqref{eq:probabilitySumBoundIof} that
\begin{align}
 \sum\limits_{l=1}
^n\sum\limits_{\vecx\in\suppX^ {(l)}} e^
{-\wfunc(\vecx)[\cprob-\epsilon]}\overset{n\rightarrow\infty}{
\longrightarrow}\infty.
\end{align}
For every $\epsilon>0$, the generating function $\gf_\dncA(s)$ thus diverges for
$\repart{s}\leq\cprob-\epsilon$. It remains to show that it converges whenever
$\repart{s}>\cprob$. For some arbitrary but
fixed $\epsilon_0>0$, define
\begin{align}
D=\sum\limits_{\{l\in\mysetN\vert \cprob+\epsilon_0 \leq
R_l\}}\sum\limits_{\vecx\in\suppX^ {(l)}}& e^
{-\wfunc(\vecx)[\cprob+\epsilon_0]}\label{eq:definitionD}
\end{align}
Because of \eqref{eq:boundAev}, the sum is taken over a finite number of terms,
and as a result, $D$ is a finite number. For every $\epsilon$ with
$\epsilon_0>\epsilon>0$, we have
\begin{align}
\sum\limits_{l=1}^n\sum\limits_{\vecx\in\suppX^ {(l)}}& e^
{-\wfunc(\vecx)[\cprob+2\epsilon]}=
\sum\limits_{l=1}^n\sum\limits_{\vecx\in\suppX^{(l)}}
e^{-\wfunc(\vecx)\epsilon}e^{-\wfunc(\vecx)[\cprob+\epsilon]}\\
&\leq
\sum\limits_{l=1}^n \sum\limits_{
\vecx\in\suppX^{(l)}}e^{-\weight[l]\epsilon}e^{-\wfunc(\vecx)[\cprob+\epsilon]}
\label{eq:boundNotTooDense}\\
&=
\sum\limits_{l=1}^n e^{-\weight[l]\epsilon}\sum\limits_{
\vecx\in\suppX^{(l)}}e^{-\wfunc(\vecx)[\cprob+\epsilon]}\\
&\leq
\sum\limits_{l=1}^n e^{-\weight[l]\epsilon}\sum\limits_{
\vecx\in\suppX^ {(l)}}e^{-\wfunc(\vecx)R_l}+D\label{eq:boundD}\\
&=\sum\limits_{l=1}^n e^{-\weight[l]\epsilon}+D.\label{eq:upperBound}
\end{align}
The inequality in \eqref{eq:boundNotTooDense} holds because for every
$l\in\mysetN$, the weight of $\vecx\in\suppXl$ is lower bounded by
$\wfunc(\vecx)\geq\weight[l]$. We have inequality in \eqref{eq:boundD}, because
of $\exp(-\weight[l]\epsilon)<1$ and \eqref{eq:boundAev}. For those $l$ for
which $R_l\leq \cprob+\epsilon$ does not apply, we add the correcting value $D$
as defined in \eqref{eq:definitionD}. We can now write the sum in
\eqref{eq:upperBound} as
\begin{align}
\sum\limits_{l=1}^n e^{-\weight[l]\epsilon}=\sum\limits_{l=1}^n
(e^{-\epsilon})^{\weight[l]}.
\end{align}
For $n$ tending to infinity, according to Lemma
\ref{lem:notTooDenseConvergence}, this series converges,
since $\{\weight[k]\}_l^\infty$ is not too dense and since $\exp(-\epsilon)<1$.
We conclude that $\gf_\dncA(s)$ converges for $\repart{s}\geq\cprob+2\epsilon$.

If, for every $\epsilon>0$, $\gf_\dncA(s)$ diverges for
$\repart{s}\leq\cprob-\epsilon$
and converges for $\repart{s}\geq\cprob+\epsilon$, then the region of
convergence of
$\gf_\dncA(s)$ is given by $\repart{s}>\cprob$. This concludes the proof of the
theorem.
\end{proof}

%% file: conclusions.tex
\section{CONCLUSIONS}
In this work, we showed that the equality of the combinatorial capacity and
the maximum entropy rate of an input process holds for constrained systems in
general and is not a consequence of regular constraints, which were considered
in this context until now. In contrast to the proof of \cite[Theorem
8]{Shannon1948} in \cite{Khandekar2000} for the regular case, our proof for the
general case is not constructive, so it
remains a challenge to explicitly define capacity achieving input sources for
constrained systems with non-regular constraints as the one considered in
\cite{Milenkovic2007}.

%% file: acknowledgement.tex
\section*{ACKNOWLEDGMENT}
We want to thank Tobias Koch for his comments on a former version of this 
paper and we would also like to thank the anonymous referees for their reviews.
Both helped substantially to improve the presentation of the material.